\documentclass{ws-procs9x6}

\begin{document}

\title{Precise measurement of \\
$\sin^22\theta_{13}$ \\
using Japanese Reactors \footnote{\uppercase{T}he theoretical working group members are \uppercase{H.M}inakata,
\uppercase{O.Y}asuda and \uppercase{H.S}ugiyama; \uppercase{D}ept. of \uppercase{P}hysics, \uppercase{T}okyo \uppercase{M}etropolitan \uppercase{U}niv.}}

\author{F.Suekane\footnote{speaker; suekane@awa.tohoku.ac.jp}, K.Inoue, T.Araki and K.Jongok}

\address{Research Center for Neutrino Science, \\ 
Graduate School of Science, \\
Tohoku University, \\
Sendai, 980-8578, Japan\\
http://www.awa.tohoku.ac.jp}  


\maketitle  

\abstracts{ 
After the KamLAND results, the remaining important targets in
neutrino experiments are to measure still unknown 3 basic
parameters; absolute neutrino mass scale, CP violation phase
$\delta_{CP}$ and last mixing angle $\theta_{13}$.    
The angle $\theta_{13}$ among them is expected to be measured in near
future by long baseline accelerator experiments and reactor experiments. 
In this paper, a realistic idea of high sensitivity reactor measurement of 
$\sin^22\theta_{13}$ is described. 
This experiment uses a giant nuclear power plant as the neutrino
source and three identical detectors are used to cancel detector and
neutrino flux uncertainties. 
The sensitivity reach on $\sin^22\theta_{13}$ is $0.017\sim0.026$ at
$\Delta m^2_{13} \sim 3 \times 10^{-3}eV^2$, which is five to seven
times better than the current upper limit measured by CHOOZ.
}

\section{Introduction}
  The year 2002 was a fruitful year for neutrino physics. 
The SNO group showed that the solar neutrino deficit is due to neutrino
transformation\cite{SNO}. 
The KamLAND group observed large deficit in reactor neutrinos and
excluded all the solar neutrino solutions except for LMA\cite{KamLAND}. 
K2K group confirmed\cite{K2K} SuperKamiokande (SK) results of atmospheric
neutrino oscillation\cite{SK}.
From these observations, four out of seven elementary parameters of neutrinos
have been measured. 
The measurements of remaining parameters, such as mixing angle
$\theta_{13}$, absolute scale of neutrino mass and CP violating phase
$\delta_{CP}$ are the next crucial issues. 

The mixing parameter $\sin^22\theta_{13}$ can be measured by disappearance
of reactor $\bar{\nu_e}$  at energy/baseline range to be around  $\Delta
m^2_{13}$, as shown below. 
\begin{equation}
P(\bar{\nu_e} \rightarrow \bar{\nu_e})=1-\sin^22\theta_{13}\sin^2\frac{\Delta m^2_{13}L}{4E_{\nu}}
\label{eq:Reactor}
\end{equation}
    
The current upper limit of $\sin^22\theta_{13}$ was measured by CHOOZ
group using reactor $\bar{\nu_e}$ to be $\leq0.12$ if $\Delta
m^2_{13}\sim3\times10^{-3} eV^2$~\cite{CHOOZ}.
In order to improve the sensitivity, a realistic idea of new-generation
reactor experiment is being investigated~\cite{minakata}. 
It uses a giant nuclear power plant of multi reactor complex as the
neutrino source.  
Identical detectors are placed at approximately oscillation maximum
baseline and near the reactors.
The data from those detectors are compared to cancel 
systematic uncertainties when extracting the disappearance rate. 
This near/far detector strategy was originally proposed by Kr2Det
group\cite{KR2DET}.  
Together with optimized baselines, detector improvements and far/near
strategy, our experiment can improve the sensitivity for
$\sin^22\theta_{13}$ significantly better than the CHOOZ experiment.

\section{Physics Motivations}
There is a number of reasons why reactor $\theta_{13}$ measurement is
important. 

(1) $\theta_{13}$ is the last neutrino mixing angle whose finite value
has not yet been measured. 
Especially it is important to know how small $\theta_{13}$ is, while
other two mixing angles are large unlike quark sector.  

(2) The size of $\theta_{13}$ is related to the detectability of leptonic
 $\delta_{CP}$ in future long baseline (LBL) accelerator
 experiments.  
The sensitivity to $\delta_{CP}$ changes rapidly at around
 $\sin^22\theta_{13}\sim0.02$ and the knowledge of $\sin^22\theta_{13}$
 down to this range will give important guideline to make strategies for
 future LBL $\delta_{CP}$ experiments~\cite{JHF}. 

\begin{figure}[ht]
\centerline{\epsfxsize=3in\epsfbox{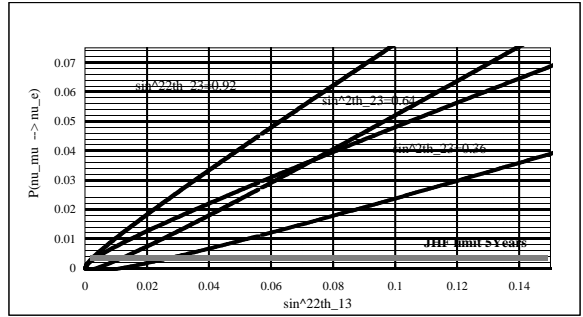}}   
\caption{
The relation between $\sin^22\theta_{13}$ and $\nu_{\mu} \rightarrow \nu_e$ 
appearance probability at $\Delta m^2_{23}$ energy-distance range.   
The matter effect is ignored for simplicity. Two bands which corresponds
 to two $\theta_{23}$ solutions for $\sin^22\theta_{23}=0.92$ are
 displayed. 
The width of each band is due to unknown $\sin\delta_{CP}$. 
Even if appearance probability is precisely measured, there are
 intrinsic ambiguities on $\sin^22\theta_{13}$.   
\label{fig:LBL}}
\end{figure}

(3) The reactor measurement of $\sin^22\theta_{13}$ is complementary
measurement to LBL $\sin^22\theta_{13}$ experiment which measures 
$\nu_e$ appearance probability in $\nu_{\mu}$ beams;
$P(\nu_{\mu}\rightarrow \nu_e)$~\cite{minakata}. 
The probability at $E_{\nu}/L=\Delta m^2_{23}/2\pi$
 is expressed in eq.(\ref{eq:LBL}).   
(For simplicity, the matter effect is ignored.) 
\begin{equation}
 \begin{split}
P(\nu_{\mu} \rightarrow \nu_e)\approx&\sin^22\theta_{13}\sin^2\theta_{23}  \\
&-\frac{\pi}{2}\frac{\Delta m^2_{12}}{\Delta m^2_{23}}\cos\theta_{13}
\sin2\theta_{12}\sin2\theta_{23}\sin2\theta_{13}\sin\delta_{CP} 
 \end{split}
\label{eq:LBL}
\end{equation}
Using best fit oscillation parameters measured by SK and
KamLAND, the coefficient for
$\sin2\theta_{13}\sin\delta_{CP}$ in the 2nd term is calculated to be
around 0.04.   
Because $\sin\delta_{CP}$ is totally unknown, the 2nd term becomes full
ambiguity when determining $\sin^22\theta_{13}$.    
Moreover, there is degeneracy of $\theta_{23}$.
That is, even if $\sin^22\theta_{23}$ is determined by $P(\nu_{\mu} \rightarrow
\nu_{\mu})$ measurements, there are two solutions for $\sin^2\theta_{23}$  if
$\sin^22\theta_{23}$ is not unity.  
Namely, if $\sin^22\theta_{23}=0.92$, which is the current lower limit from SK,
$\sin^2\theta_{23}=0.64$ or 0.36. 
These circumstances are described in detail in the
references\cite{8degeneracy}. 
Fig.-\ref{fig:LBL} shows the relation of the appearance probability and
$\sin^22\theta_{13}$, taking into account these ambiguities. 
The sensitivity of the JHF experiment, for 
example, on $\sin^22\theta_{13}$ is limited to $\approx$0.025 ($\approx$0.015)
depending upon 
possible $\theta_{23}$ degeneracy is (is not) taken into
account~\cite{sugiyama}. 

On the other hand, the reactor experiment is pure $\sin^22\theta_{13}$
measurement and by combining reactor data and LBL data, there is a
possibility to resolve ambiguities of $\theta_{23}$ degeneracy and
$\Delta m^2_{23}$ hierarchy and even access to
$\sin\delta_{CP}$\cite{minakata}.  
\section{The Experiment}
\begin{figure}[ht]
\centerline{\epsfxsize=2in\epsfbox{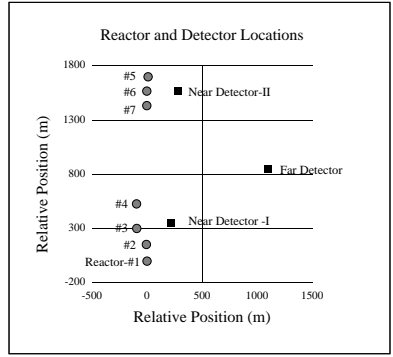}}   
\caption{Reactor (circles) and detector (squares) relative locations. 
There are 7 reactors in Kashiwazaki-Kariwa nuclear power plant,
 producing 24.3GW maximum thermal power. 
Reactor \#1 through \#4 form a cluster and \#5 through \#7 form another
 cluster.  
The two clusters separate about 1.3km apart. 
Two near detectors will be placed at around 300 to 350m from each cluster. 
The far detector will be placed at around 1.3km from all the reactors. 
\label{fig:kash}}
\end{figure}

\begin{figure}[ht]
\centerline{\epsfxsize=3.6in\epsfbox{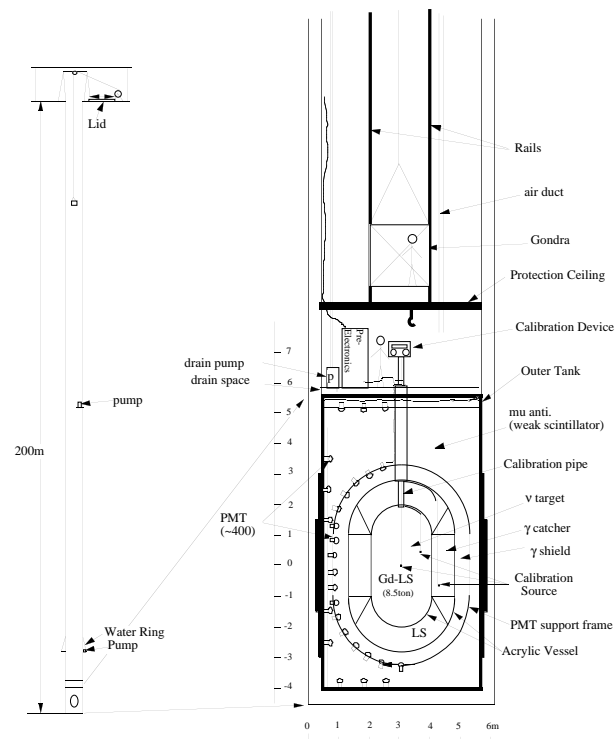}}
\caption{Schematic view of the detector. 
The $\bar{\nu_e}$ target is 8.5ton Gd loaded liquid scintillator. 
The $\bar{\nu_e}$ target is surrounded by 70cm thick $\gamma$ catcher
 scintillator.
The $\gamma$ catcher scintillator is surrounded by 60cm thick buffer
 scintillator with very slight light output.
The outer most layer is muon anti-counter made of the same scintillator
 as the buffer region. 
The far detector will be placed at the bottom of 200m shaft hole with
 diameter 6m. 
The near detectors will be placed at the bottom of 70m depth shaft hole.   
\label{fig:detector}}
\end{figure}

In this experiment, three identical detectors are build in the site of
Kashiwazaki-Kariwa nuclear power plant (NPP) which is operated by Tokyo
Electric Power Company. 
The Kashiwazaki NPP has 7 reactors, producing total thermal energy of 24.3GW.
This is the most powerful NPP in the world.
 Using large-power  nuclear power plant is profitable for not only
obtaining high event rate but also realizing low background to signal ratio
at a given depth underground.
The relative locations of reactors and detectors
are shown in the fig.-\ref{fig:kash}.
Although, the far/near distance ratios between the reactors and
detectors are not unique, the uncertainty introduced from the variations
of the distances are estimated to be only 0.2\%.
The detector is CHOOZ like detector as shown in the
fig.-\ref{fig:detector}.  
The central part; the $\bar{\nu_e}$ target is 8.5 ton Gadolinium loaded liquid
scintillator. 
  The component of the liquid scintillator is the PaloVerde type, which
was proven to be stable in an acrylic container\cite{PaloVerde}. 
The Gd concentration is 0.15\% which is 1.5 times  higher than that of
CHOOZ scintillator. 
The higher Gd concentration is intended to increase the neutron
absorption efficiency on Gd and to reduce the systematic uncertainty
associating with the inefficiency. 
Our preliminary study shows that the scintillator is stable with 0.15\%
Gd concentration.   
The reactor $\bar{\nu_e}$ is detected by the following inverse $\beta$
decay reaction.
\begin{equation}
\bar{\nu_e} + p \rightarrow e^+ + n
\end{equation}
The positron annihilates with electron within a few nano seconds
after slowing down in the scintillator material, then produces two
0.511MeV $\gamma$'s.  
These process produces a prompt signal, whose energy is between 1MeV
and 8MeV.   
On the other hand, the produced neutron is thermalized quickly and
absorbed by Gd, producing $\gamma$ rays whose total energy
amounts to 8MeV. 
The neutron absorption occurs typically 20$\mu s$ after the prompt signal.  
By requiring the timing correlations between the positron signal and
the neutron signal, backgrounds can be severely suppressed.  
The 2nd layer is unloaded liquid
scintillator whose light output is adjusted to be the same as the
$\bar{\nu_e}$-target scintillator.   
This layer works as $\gamma$-ray catcher. 
When neutrino events occur near the detector edge,
$\gamma$-rays from positron annihilation and neutron absorption may
escape from the detector. 
The $\gamma$-ray catcher is used to catch such $\gamma$-rays and to reconstruct
the original energy. 
The energy threshold for prompt signal is set to be below minimum
positron energy (1.022MeV).  
In this way no systematic ambiguities associated with threshold cut is
introduced.
The fiducial volume is defined by the existence of correlated signals.
That is, when 8MeV of energy deposit is observed after associating
prompt signal whose energy is greater than 1MeV, this event is
considered to be $\bar{\nu_e}$ event, regardless the positions of
prompt and delayed signals. 
As no position cut is necessary, this method is free from position
reconstruction error.  
The total volume of the liquid scintillator in the acrylic vessel can be
measured precisely from the liquid level in the thin calibration pipe
even if there is a distortion of the vessel after the installation.  
The 3rd and 4th layers are also liquid scintillator which has a very
slight scintillation light output. 
These layers work as a shield of gamma rays and as cosmic ray
anti-counters.  
The slight light output is to detect low energy muons whose velocity is
below the Cherenkov threshold. 
Intense calibration work will be essential in this experiment to monitor
the detector condition change. 
The whole detector will be placed in the bottom of the shaft hole with
6m diameter and 200m depth (far detector) and 70m depth (near detectors).
Digging such shaft holes can be done using existing 6m diameter
vertical drilling machine.
The background rate is expected to be less
than 2\%.  
The major component of the background comes from fast neutrons produced
in nuclear interaction caused by cosmic rays going through the rock near by.
The visible energy distribution of the prompt signal in the fast neutron
backgrounds was measured to be flat by CHOOZ group at the energy range
below 30MeV\cite{CHOOZ} and this kind of background rate can be
estimated by using the event rate within non-reactor-$\bar{\nu_e}$
energy range, such as below 1MeV and beyond 10MeV. 

The systematic error in CHOOZ experiment was 1.7\% (detector associated)
$\oplus$ 2.1\%(neutrino flux associated).    
By improving the detector system as described above, the detector
associated systematics will reduce to 1.1\%.  
When the far and near detectors are compared the ambiguity in neutrino
flux mostly  cancels. 
Even without cancellation of the detector associate systematics, the
systematic error at this stage can be reduced to 1.1\%. 
Prediction of how good the detector systematics cancellation will be is
difficult.  
However, in Bugey case, their systematic error reduced to be half of the
original error after comparing three detectors~\cite{Bugey}.  
If the same ratio is applied to this case,  the detector associated
systematics is expected to be reduced to 0.5\% after taking near/far ratio. 
All these considerations are summarized in the table-\ref{tab1} and the
the total systematic error will be 0.5$\sim$1\%, where $\sim$1\% is for the
case that the detector cancellation does not work so well.

\begin{figure}[ht]
\centerline{\epsfxsize=1.9in\epsfbox{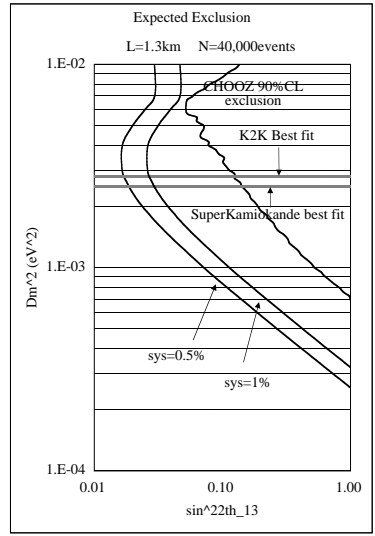}}   
\caption{The expected 90\%CL exclusion region of this experiment for the
 case of $\sigma_{sys}$=1\% and 0.5\% obtained by rate only analysis.  
At  $\Delta m^2\sim3\times 10^{-3}eV^2$, $\sin^22\theta_{13}<0.026$ and
 $<0.017$ are possible, respectively. 
\label{fig:sensitivity}}
\end{figure}

In two years of operation, 40,000 neutrino events will be recorded in far
detector and ten times more in each near detector.  
The statistic error will be 0.5\%.
The 90\% CL sensitivity of this experiment is shown in the
fig.-\ref{fig:sensitivity}.
At $\Delta m^2 \sim 3\times10^{-3} eV^2$, the sensitivity of
0.017$\sim$0.026 is expected. 
This is five to seven times better limit than CHOOZ and comparable to the LBL
sensitivity on $\sin^22\theta_{13}$.  
\begin{table}[ph]
\tbl{systematic error (\%) at each stage.}
{\footnotesize
\begin{tabular}{|c|r|r|r|}
\hline
{} &{} &{} &{} \\[-1.5ex]
{} & Detector & Flux & Total \\[1ex]
\hline
{} &{} &{} &{} \\[-1.5ex]
(1) Original (CHOOZ)                         &1.7        &2.1 &2.7\\[1ex]
(2) Detector Improvement                     &1.1        &2.1 &2.4\\[1ex]
(3) Far/Near $\bar{\nu_e}$ flux Cancellation &1.1        &0.2 &1.1\\[1ex]
(4) Far/Near Detector Cancellation         &0.5$\sim$1 &0.2 &0.5$\sim$1\\[1ex] 
\hline
\end{tabular}\label{tab1} }
\vspace*{-13pt}
\end{table}
\section{Summary and Discussions}
The reactor measurements of $\sin^22\theta_{13}$ is important because it
is a pure  $\sin^22\theta_{13}$ measurement and plays
complimentary role to LBL experiments. 
The Kashiwazaki experiment is realistic.
By comparing 3 improved CHOOZ like detectors placed at appropriate
locations from multi reactors, it is possible to measure
$\sin^22\theta_{13}$ down to 0.017 to 0.026. 
If this experiment observe positive result, the accessibility to
$\sin\delta_{CP}$ is high for future LBL experiment. 
Also there is a chance to determine $\theta_{23}$ degeneracy,
$\Delta m^2_{23}$ hierarchy, by combining with LBL data, and even to obtain a
clue to nonzero $\sin\delta_{CP}$ before going to $\bar{\nu}$ mode.  
If this experiment observes negative result, it means that $\nu_3$
component in $\nu_e$ is very small, while all other
components are hundred times larger. 
This peculiar fact may become a key information to build unified theory of
elementary particles.  
\section*{Acknowledgments}
FS thanks to Prof. A.Piepke for providing a sample of Gd loaded liquid
scintillator and giving us precious advice of the treatment. 
FS thanks to Prof. H. de Karret for useful discussions about CHOOZ experiment.


\begin{thebibliography}{0}
\bibitem{SNO} SNO Collaboration, Phys. Rev. Lett. vol.89, 011301(2002).
\bibitem{KamLAND} KamLAND Collaboration, Phys. Rev. Lett. vol.90,
	021802-1 (2002).
\bibitem{K2K} K2K Collaboration, Phys. Rev. Lett., vol.90, 041801(2003).
\bibitem{SK} Super-Kamiokande Collaboration, Phys. Rev. Lett. 81, 1562
	(1998); Phys. Rev. Lett. 85, 3999(2000).
\bibitem{CHOOZ} CHOOZ Collaboration, Eur.Phys.J. C27(2003) 331-374.
\bibitem{minakata} H.Minakata, H.Sugiyama, O.Yasuda, K.Inoue and F.Suekane,
	hep-ph/0211111.
\bibitem{KR2DET} V.Martemianov et al., hep-ex/0211070.
\bibitem{JHF} For example, JHF LoI, (http://neutrino.kek.jp/jhfnu/).
\bibitem{8degeneracy} J. Burguer-Castell et al., Nucl. Phys. B 608, 301
	(2001); H.Minakata et al., JHEP0110, 001(2001);
	H.Minakata et al., Nucl.Phys. Proc. Suppl. 110, 404(2002);
	V.Barger et al., Phys. Rev. D65, 073023(2002).
\bibitem{sugiyama}  H. Sugiyama, Talk at NuFACT03, Columbia University,
	June 5-11, 2003. For similar estimate, see Huber et al.,
	hep-ph/0303232. 
\bibitem{PaloVerde} F.Bohem et al., hep-ex/0003022.
\bibitem{Bugey} Achkar et al., Nucl. Phys. B434, 503-534 (1995).
\end{thebibliography}
\end{document}